\documentclass{PoS}
\newcommand{\pptopp}{\nobreak{pp\to W^+W^+jj}}
\newcommand{\pptopm}{\nobreak{pp\to W^+W^-jj}}
\newcommand{\WpWp}{\nobreak{W^+W^+jj}}
\newcommand{\WpWm}{\nobreak{W^+W^-jj}}
\title{Weak Bosons and Jets at the LHC}

\ShortTitle{Weak Bosons and Jets at the LHC}

\author{\speaker{Tom Melia}\thanks{Based on work done in collaboration with Kirill Melnikov, Paolo Nason, Raoul R\"ontsch, and Giulia Zanderighi - see Refs.~\cite{Melia:2010bm,Melia:2011gk,Melia:2011dw}.}\\
        University of Oxford Theoretical Physics\\
        E-mail: \email{t.melia1@physics.ox.ac.uk}}

\abstract{In this talk, I outline theoretical predictions for weak boson pair production 
in association with two jets at the LHC. I will discuss the next-to-leading
order QCD corrections to the processes $\pptopp$ and $\pptopm$, and the interfacing
of $\pptopp$ with a parton shower using the {\tt POWHEG BOX} framework. }

\FullConference{ 10th International Symposium on Radiative Corrections (Applications of Quantum Field Theory to Phenomenology) - Radcor2011\\
September 26-30, 2011\\
Mamallapuram, India}

\begin{document}

\section{Introduction and Motivation}

\begin{figure}[t]
\begin{center}
\includegraphics[angle=0,scale=1.40]{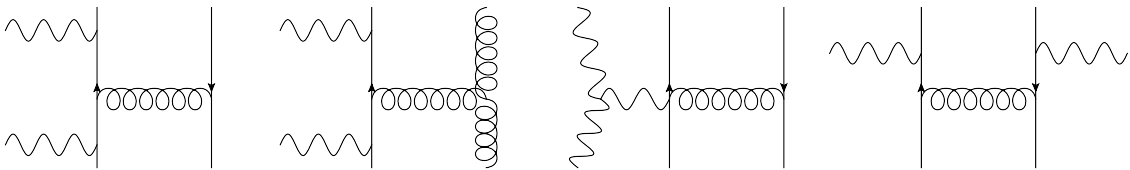}

\caption{Types of Feynman graphs encountered in a tree-level calculation of weak boson pair production in association with two jets. 
Wavy lines depict weak bosons and spiralled lines are gluons. Quark flavour labels and weak boson labels ($W^+/W^-/Z^0$) have been 
deliberately suppressed to highlight the general topology of the graphs which may contribute to different processes.}
\label{feyn}
\end{center}
\end{figure}

Precision calculations of standard model processes are essential for interpreting the signals measured at the Large Hadron Collider (LHC) and for
fully realising the discovery potential of this experiment. Next-to-leading order (NLO) calculations in perturbative QCD have proved very successful when used 
in Tevatron analyses, and are a good way to reduce theoretical uncertainties in a description of a given process. On top of this, merging an 
NLO calculation with a parton shower provides a realistic hadron-level prediction for an event whilst maintaining NLO accuracy for inclusive observables.

In this talk I shall discuss the production of a pair of weak bosons in association with jets -- specifically the two processes $\pptopp$ and $\pptopm$.
I will describe the computation of the NLO QCD corrections to both processes, as well as the merging of $\pptopp$ with a parton shower, done in the framework
of the {\tt POWHEG BOX} \cite{Alioli:2010xd}.

Figure~\ref{feyn} depicts the structure of some of the types of Feynman graphs one encounters in a tree-level calculation of weak boson pair 
production along with two jets. All of these types of graph contribute towards the process $\pptopm$, but only graphs of the type shown in the far right of the figure
contribute towards $\pptopp$. Here, charge conservation requires the two $W^+$ bosons to be emitted from separate quark lines and
this leads to an unusual theoretical property -- the cross section for this process remains finite even if the requirement that two jets are observed is 
lifted. This will be investigated later on, and I will present results for $W^+W^+ + n$~jets, where $n=0,1,2$. The calculation of $\pptopp$
can be seen as a stepping stone to the calculation of $\pptopm$, since it involves a small subset of the Feynman graphs needed for the latter.

Both processes are $2\to4$ processes, and to calculate the QCD corrections to them one needs to deal with one-loop, six-point tensor integrals of relatively high rank. 
There is thus a theoretical incentive in performing these calculations
and much progress has been made over the past few years in the methods used to compute them -- this will be discussed in the following 
section. But before this, I will go on to discuss the study of both processes at the LHC in a bit more detail.

\subsection{$\WpWp$ at the LHC}
At $\sqrt{s}=$14~TeV, the cross-section for this process is about 1 pb (40\% of this for $W^-W^-jj$) and is therefore accessible. In the following we take
the $W^+$ bosons to both decay leptonically, giving rise to a nearly background-free signature which involves same-sign leptons. This is an interesting
process to study in its own right, but there are other reasons to study it: $\pptopp$ is a background to physics both 
within and beyond the standard model. For example, it is possible to use same-sign lepton pairs to study double parton scattering at the LHC \cite{Gaunt:2010pi}, to 
which $\pptopp$ is a background. Beyond the standard model, resonant slepton production in R-parity violating 
SUSY models \cite{Dreiner:2006sv}, diquark production \cite{Han:2009ya}, and doubly charged Higgs boson production \cite{Maalampi:2002vx} are examples of processes which also
 lead to a signature of two same-sign leptons, missing energy, and jets.

\subsection{$\WpWm$ at the LHC}
The production of a $W^+W^-$ boson pair in association with zero, one or two jets is an important background to Higgs boson production, especially when the decay $\nobreak{H\to W^+W^-}$ opens up. Although most of the sensitivity in Higgs boson searches comes from the zero jet processes, which have the largest cross-section,
the production of a Higgs boson in association with two jets is also relevant -- about 10\% of Higgs events at the LHC involve two jets 
\cite{Anastasiou:2009bt,Campbell:2006xx}. 
The production of a Higgs boson via weak boson fusion (WBF) also has a sizeable cross-section. The signature of this process includes two forward tagging jets 
and $\pptopm$ is an irreducible background to this. As we did for $\WpWp$, in the following we will take both $W$ bosons to decay leptonically. 
The resulting signature of two opposite-sign leptons, jets and missing energy is also a background to a classic beyond the standard model physics search.

\section{Method of calculation}

\subsection{The NLO QCD corrections}
NLO QCD calculations of processes involving more than five particles is difficult. For the virtual amplitude, the number of Feynman diagrams needing evaluation
 grows factorially with the number of particles in the process. In addition to this, the one-loop tensor integrals which need to be computed become more involved. 
However, a refinement of traditional computation methods, as well as the development of new techniques based on unitarity and on-shell methods, 
have seen a significant growth in the
number of $2\to4$ processes (and even a $2\to5$ process) known at NLO in the past few years (see \cite{Ellis:2011cr} for a recent review). 
Platforms for the automation of NLO-accurate processes are
currently being developed (see e.g. \cite{Cullen:2011ac, Bevilacqua:2011xh, Hirschi:2011pa, Berger:2008sj}).

As described in detail in the papers \cite{Melia:2010bm,Melia:2011dw}, the technique of $D$-dimensional generalised unitarity \cite{Giele:2008ve} was 
used to obtain the virtual part of the amplitude for the QCD processes $\pptopp$ and $\pptopm$. It is worth pointing out that, as currently formulated, 
on-shell methods require working with an ordering of external lines -- these are colour ordered or primitive amplitudes. It is only colour-charged
particles which are ordered in primitive amplitudes and so all possible insertions of the colourless weak bosons must be considered for any tree-level or one-loop
primitive amplitude. The $D$-dimensional unitarity cuts reduce one-loop primitive amplitudes to products of tree-level helicity amplitudes, and a certain amount of 
difficulty exists in ensuring no over-counting takes place when combining the cuts of different parent diagrams. Nevertheless, this is just book-keeping and these
two calculations demonstrated
that unitarity methods can deal with more complicated, colourless final states. The tree-level helicity amplitudes themselves are calculated using Berend-Giele recursion
relations \cite{Berends:1987me}.

\subsection{Merging with a parton shower}

Methods which include both the benefits of an NLO calculation and a parton shower model (NLO+PS generators) have become available in recent years 
- two frameworks  are currently being used for collider physics:  {\tt MC@NLO} \cite{Frixione:2002ik} and {\tt POWHEG} \cite{Nason:2004rx}.
A general computer framework for building a {\tt POWHEG} implementation of an arbitrary NLO process exists 
- the {\tt POWHEG BOX} \cite{Alioli:2010xd}. Here, one needs only to supply a few ingredients: phase-space and flavour information, the Born and real matrix elements and the
virtual matrix elements for a given NLO process. The implementation of $\pptopp$ in the {\tt POWHEG BOX} is reported in \cite{Melia:2011gk}. This was the first
time a $2\to4$ process was implemented in a NLO+PS generator. 

Since all of the ingredients needed by the {\tt POWHEG BOX} were already known from \cite{Melia:2010bm}, the {\tt POWHEG} implementation of this process 
did not present any special problem, except for a non-trivial issue of high computational demands coming from the virtual corrections. 
The technical details of how this problem was dealt with are described in detail in \cite{Melia:2011gk}. The resulting code is public and is available at the website 
\cite{website}.

\section{Results}

\begin{figure}[t]
\begin{center}
\includegraphics[angle=0,scale=0.35]{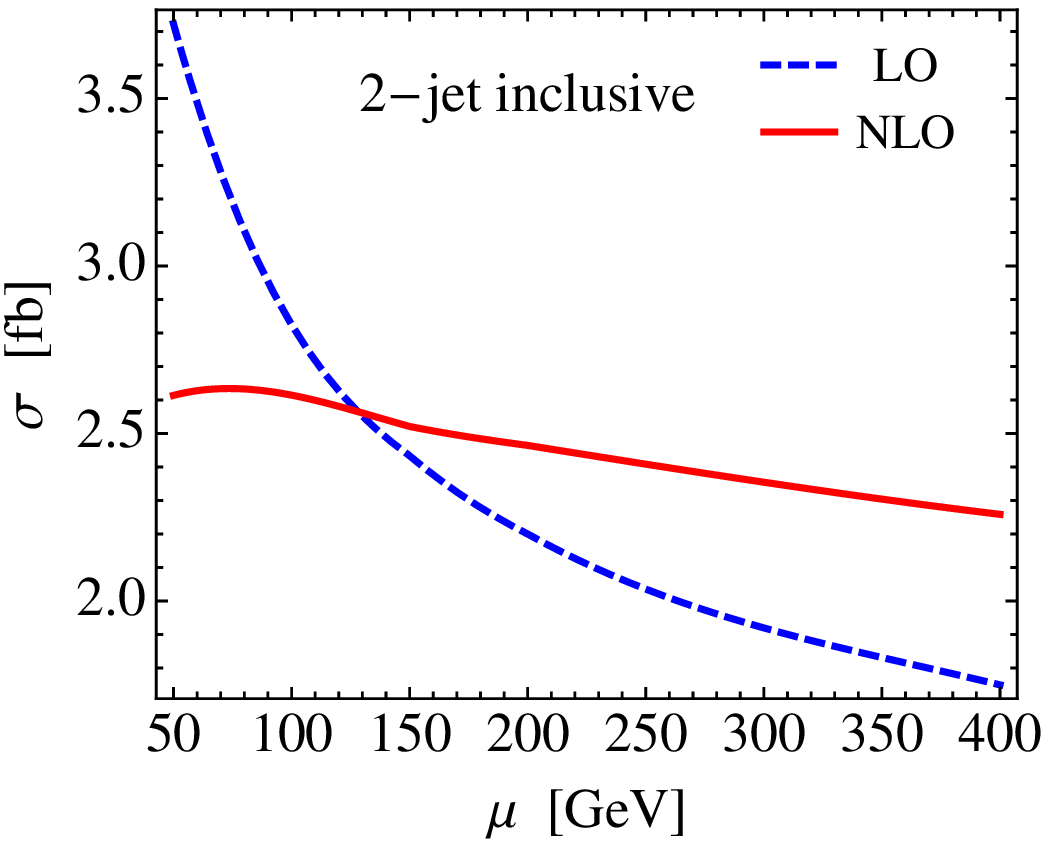}
\includegraphics[angle=0,scale=0.33]{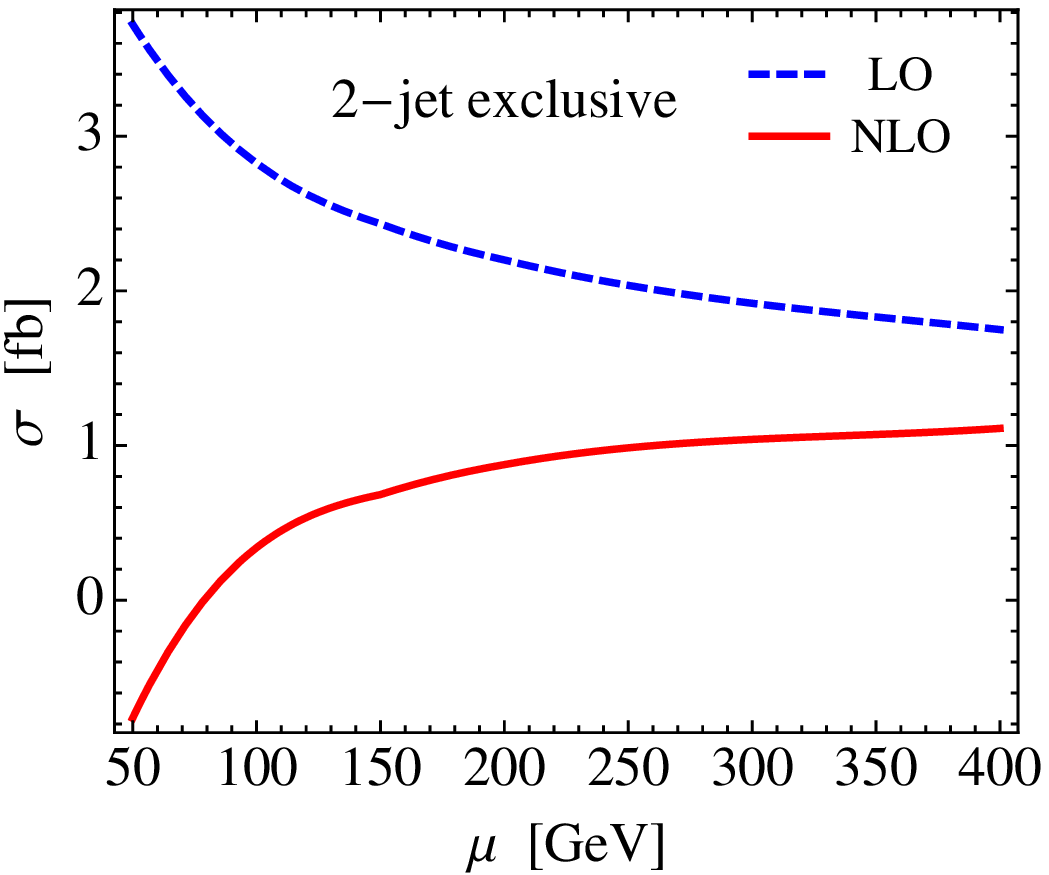}
\includegraphics[angle=0,scale=0.35]{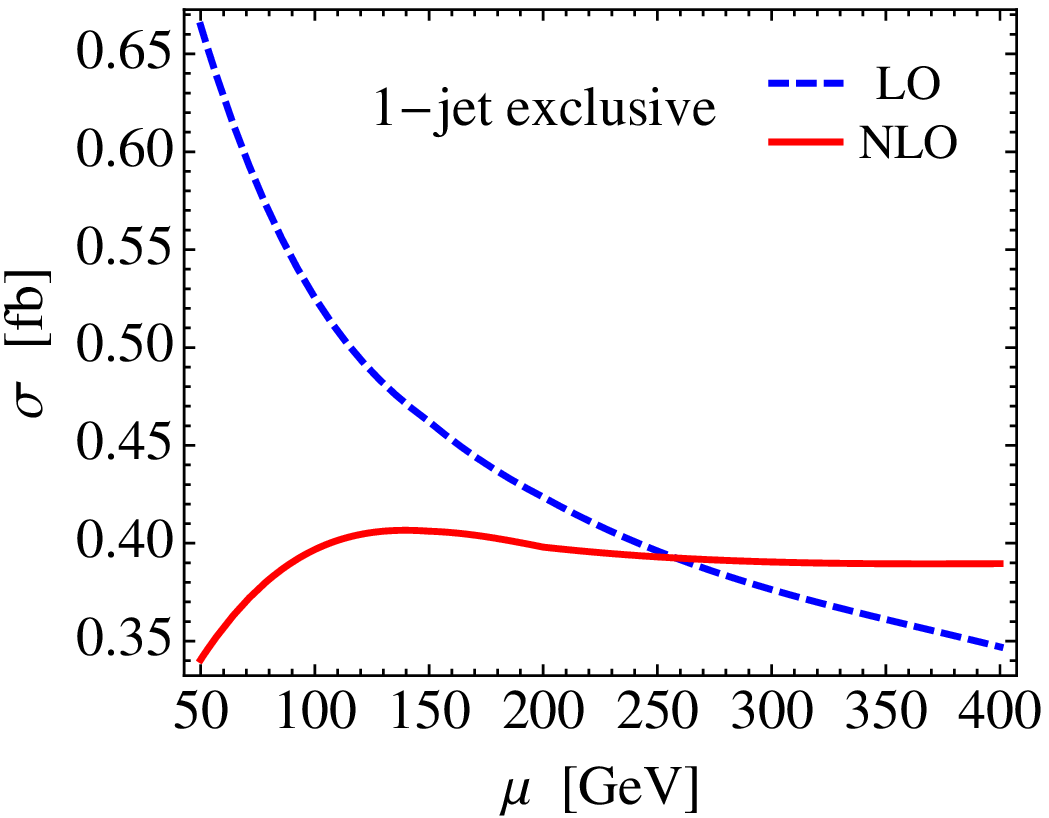}
\includegraphics[angle=0,scale=0.35]{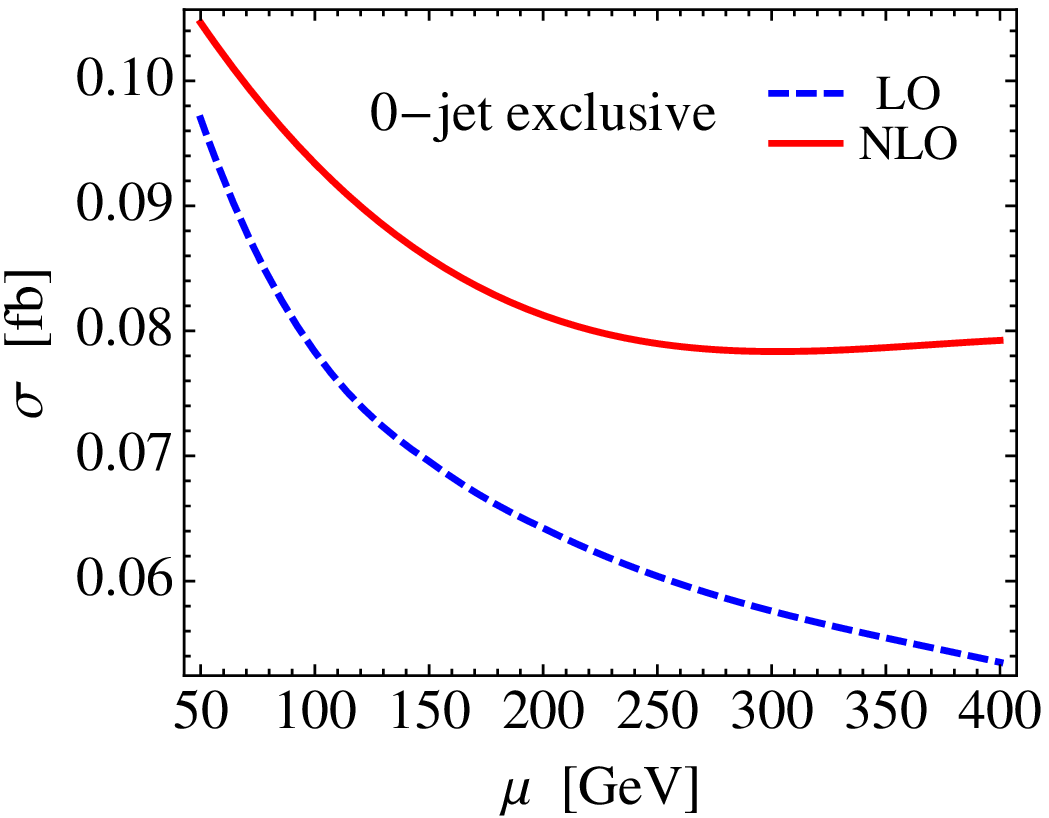}
\caption{The dependence on factorisation and renormalisation scales of
  cross-sections for $\nobreak{pp \to e^+\, \mu^+\, {\nu}_{e}\, {\nu}_{\mu} + n~{\rm
    jets}}$, $n = 0,1,2$ at leading and next-to-leading order in
  perturbative QCD. Here $\mu_{\rm F} = \mu_{\rm R} = \mu$. }
\label{wpwpxsec}
\end{center}
\end{figure}

\subsection{Selected results for $\pptopp$ at the LHC}
First I will present results from the NLO calculation of $\pptopp$, taken from \cite{Melia:2010bm}.
We consider proton-proton collisions at a center-of-mass energy
$\sqrt{s} = 14~{\rm TeV}$. We require leptonic decays of the
$W$-bosons and consider the final state $e^+ \mu^+ \nu_e \nu_\mu $ . The
$W$-bosons are on the mass-shell and we neglect quark flavour mixing.
We impose standard cuts on lepton transverse momenta $p_{\perp, l} >
20~{\rm GeV}$, missing transverse momentum $p_{\perp, \rm miss} >
30~{\rm GeV}$ and charged lepton rapidity $| \eta_l| < 2.4$. We define
jets using anti-$k_{\perp}$ algorithm, with
$R = 0.4$ and with a
transverse momentum cut $p_{\perp, j} = 30~{\rm GeV}$ on the two
jets. The mass of the $W$-boson is taken to be $m_W = 80.419~{\rm
  GeV}$, the width $\Gamma_W = 2.140$~{\rm GeV}. $W$ couplings to
fermions are obtained from $\alpha_{\rm QED} (m_Z) = 1 / 128.802$ and
$\sin^2 \theta_W = 0.2222$.  We use MSTW08LO parton distribution
functions for leading order and MSTW08NLO for next-to-leading order
computations, corresponding to $\alpha_s(M_Z) = 0.13939$ and
$\alpha_s(M_Z) = 0.12018$ respectively. We do not
impose lepton isolation cuts. All results discussed below apply to the
QCD production $pp \to W^+ W^+ jj$; the electroweak contribution to
this process is ignored.  

\begin{figure}[t]
\begin{center}
\includegraphics[angle=0,scale=0.55]{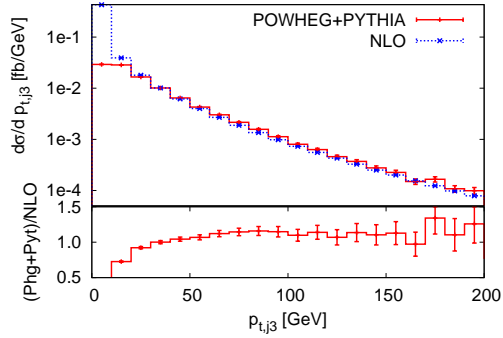}
\caption{The kinematic distribution for the transverse momentum of the third hardest jet in the QCD production of $pp\to e^+\mu^+\nu_e\nu_\mu+2~$jets. The pure 
NLO result and the result with {\tt POWHEG+PYTHIA} are both shown.}
\label{pwg}
\end{center}
\end{figure}

Figure~\ref{wpwpxsec} shows the dependence of the
production cross-sections for $pp \to e^+ \mu^+ \nu_{e} \nu_{\mu} +
n~{\rm jets}$ on the renormalisation and factorisation scales, which
we set equal to each other.
Considering the
range of scales $50~{\rm GeV} \le \mu \le 400~{\rm GeV}$, we find the
two-jet inclusive cross-section to be $\sigma^{\rm LO} = 2.7 \pm
1.0~{\rm fb}$ at leading order and $\sigma^{\rm NLO} = 2.44 \pm
0.18~{\rm fb}$ at next-to-leading order. The forty percent scale
uncertainty at leading order is reduced to less than ten percent at
NLO.  We observe similar stabilization of the scale dependence for the
$0$- and $1$-jet exclusive multiplicities.  Combining these
cross-sections we obtain a total NLO cross-section of about $2.90~{\rm
  fb}$ for $pp \to e^+ \mu^+ \nu_e \nu_\mu $ inclusive
production. This implies about $60$ $e^+\mu^++ e^+e^+ + \mu^+\mu^+$
events per year at the LHC with $10~{\rm fb}^{-1}$ annual
luminosity. While this is not a gigantic number, such events will have
a very distinct signature, so they will definitely be seen and it will
be possible to study them. 

The dramatic change in the two-jet exclusive cross-section apparent from 
 figure~\ref{wpwpxsec} is discussed and investigated in \cite{Melia:2010bm}. We find that 
 the feature observed here, that the two-jet exclusive is significantly smaller
 than the two-jet inclusive, remains present when we increase the jet cut and 
 so allow for greater perturbative convergence of the exclusive cross section. 
 This smallness implies that quite
a large fraction of events in $pp \to e^+\mu^+ \nu_e \nu_\mu + \ge
2~{\rm jets}$ have a relatively hard third jet. This feature may be
useful for rejecting contributions of $\pptopp$ when looking
for multiple parton scattering.

Next I present results from the {\tt POWHEG} implementation of $\pptopp$, taken from the paper \cite{Melia:2011gk}. Here the set-up is as described above, but
we consider $pp$ collisions at a different centre of mass energy: $\sqrt{s} = 7~{\rm TeV}$. A dynamic scale is used for the renormalisation and factorisation scales: 
\begin{eqnarray}
\mu_R=\mu_F=(p_{\perp,1}+p_{\perp,2}+E_{\perp,W_1}+E_{\perp,W_2})/2, ~~~~E_{\perp,W}= \sqrt{m_W^2+p_{\perp,W}^2}, \nonumber
\end{eqnarray}
where $p_{\perp,W_1}$, $p_{\perp,W_2}$, $p_{\perp,1}$ and $p_{\perp,2}$ are the transverse momenta of the two $W$s and the two emitted partons in the underlying Born configuration. 

With no jet cuts, but with the leptonic cuts described above, we find the cross-section for to be $1.11\pm0.01~$fb for the pure NLO result, and a slightly lower cross-section of  $1.06\pm0.01~$fb when events are generated by {\tt POWHEG} and are subsequently showered with {\tt PYTHIA}. A comparison of kinematic distributions was carried out in \cite{Melia:2011gk} and for the most part, there was good agreement between the NLO and the {\tt POWHEG+PYTHIA} results. However, there were some distributions which showed expected and marked changes, one of which I shall highlight in this talk. Figure~\ref{pwg} shows the transverse momentum of the 
third-hardest jet. Since at NLO it is only the real radiation which contributes to this distribution, we see a divergence for small $p_{\perp,j_3}$ in the pure NLO result.
In contrast one can see the Sudakov peak in the {\tt POWHEG + PYTHIA} result, and the distribution goes to zero as $p_{\perp,j_3}\to0$.

\subsection{Selected results for $\pptopm$ at the LHC}

\begin{figure}[t]
\begin{center}
\includegraphics[angle=0,scale=0.55]{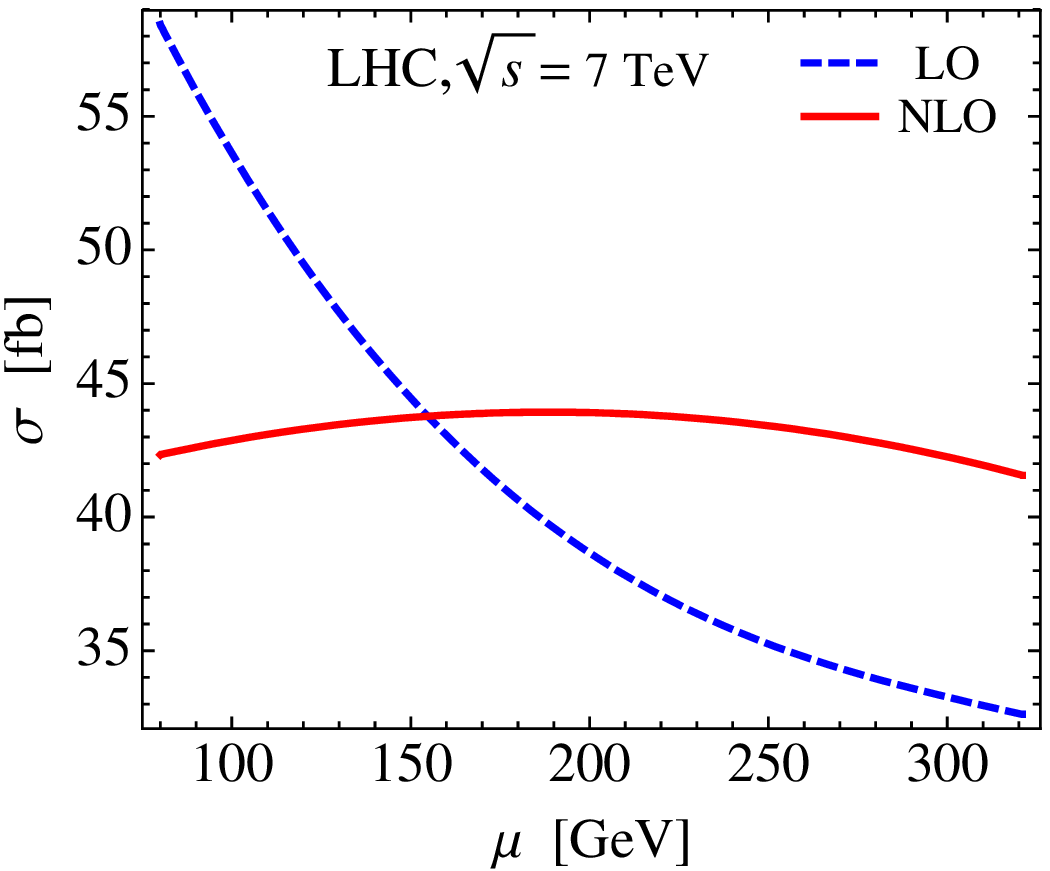}
\includegraphics[angle=0,scale=0.55]{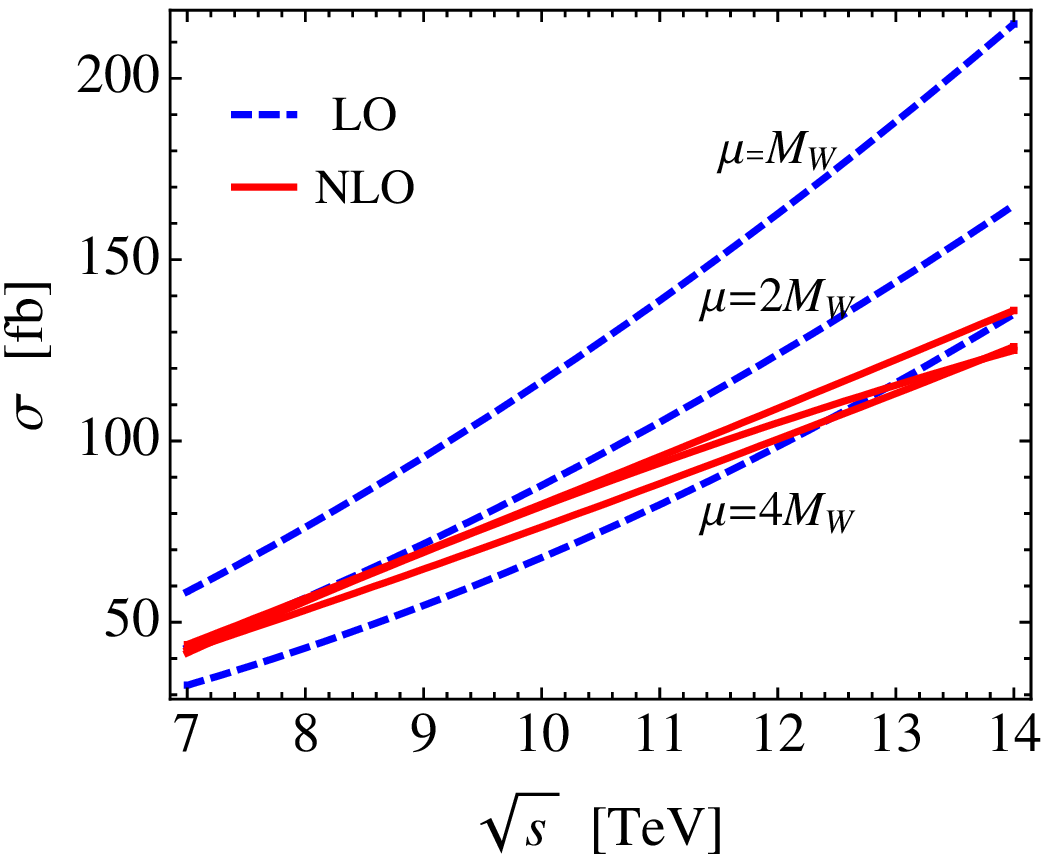}
\caption{Left pane: the production cross-section of the process $pp\to(W^+\to\nu_ee^+)(W^-\to \mu^-\bar{\nu}_\mu)jj$ at the $7~$TeV LHC in dependence on the 
factorisation and renormalisation scales $\mu_F=\mu_R=\mu$ at both LO and NLO in perturbative QCD. Right pane: the dependence of the cross-section
on centre of mass energy $\sqrt{s}$ with LO results in dashed blue and NLO results in solid red. Three choices of $\mu$ are shown: $\mu=m_W,2m_w,4m_W$. }
\label{wpwmxsec}
\end{center}
\end{figure}

Here I will present selected results from the calculation of $\pptopm$, taken from the paper \cite{Melia:2011dw}. Here the $W$ bosons decay leptonically: $\nobreak{W^+W^-jj\to e^+\mu^-\nu_e\bar{\nu}_\mu jj}$. The full results with generic opposite-sign leptons can be obtained from these by multiplying by a factor of four. We use the
same leptonic cuts and electroweak input parameters as were described in the $\pptopp$ results section above. However, here of course a jet cut must be applied
and two jets observed in order to obtain a finite cross-section: we take $p_{\perp,j}>30~$GeV and $|\eta_j|<3.2$. 

Figure~\ref{wpwmxsec} shows the dependence of the production cross-section on renormalisation and factorisation scales, which are again set equal to each other, 
at a centre of mass energy $\sqrt{s} = 7~{\rm TeV}$. The dependence of the cross-section on centre of mass energy is also shown in figure~\ref{wpwmxsec}. One observes
a dramatic reduction in scale dependence in going from leading order to next-to-leading order. Considering a range of scales $m_W<\mu<4m_W$ we obtain
a cross section at leading order $\sigma_{LO}=46\pm13~$fb and at NLO $\sigma_{NLO}=42\pm1~$fb. Assuming fifty percent efficiency,
with 5~fb$^{-1}$ of data at the 7~TeV run of the LHC, we expect about 400 dilepton events $e^+\mu^-$,$e^+e^-$,$\mu^+e^-$,$\mu^+\mu^-$. It is interesting that
at NLO, the dependence of the cross-section on centre of mass energy $\sqrt{s}$ is almost linear. If one defines an `optimal' scale choice to be the choice
of scale for which NLO corrections are smallest then this `optimal' scale shifts from $2m_W$ at $7~$TeV to $4m_W$ at $14~$TeV.

\begin{figure}[t]
\begin{center}
\includegraphics[angle=0,scale=0.55]{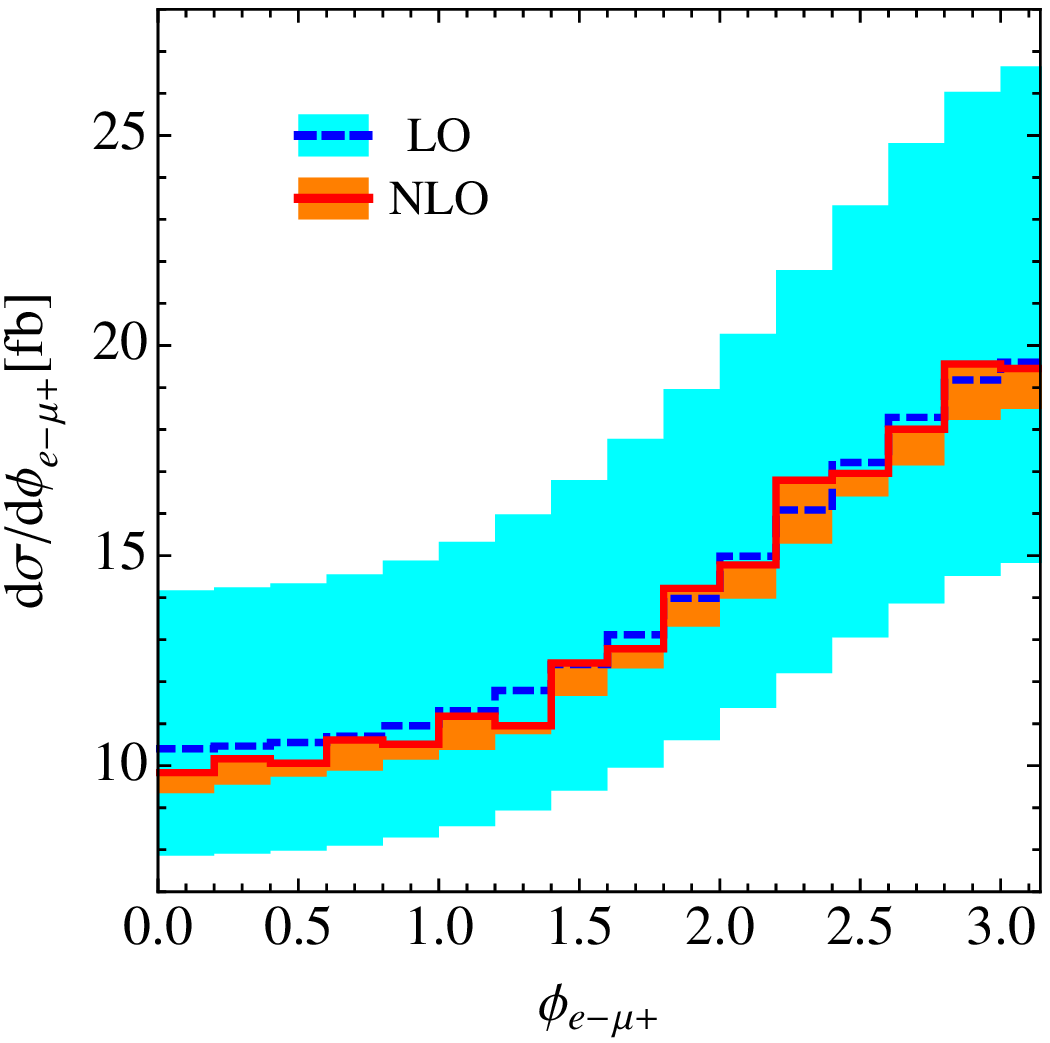}
\includegraphics[angle=0,scale=0.55]{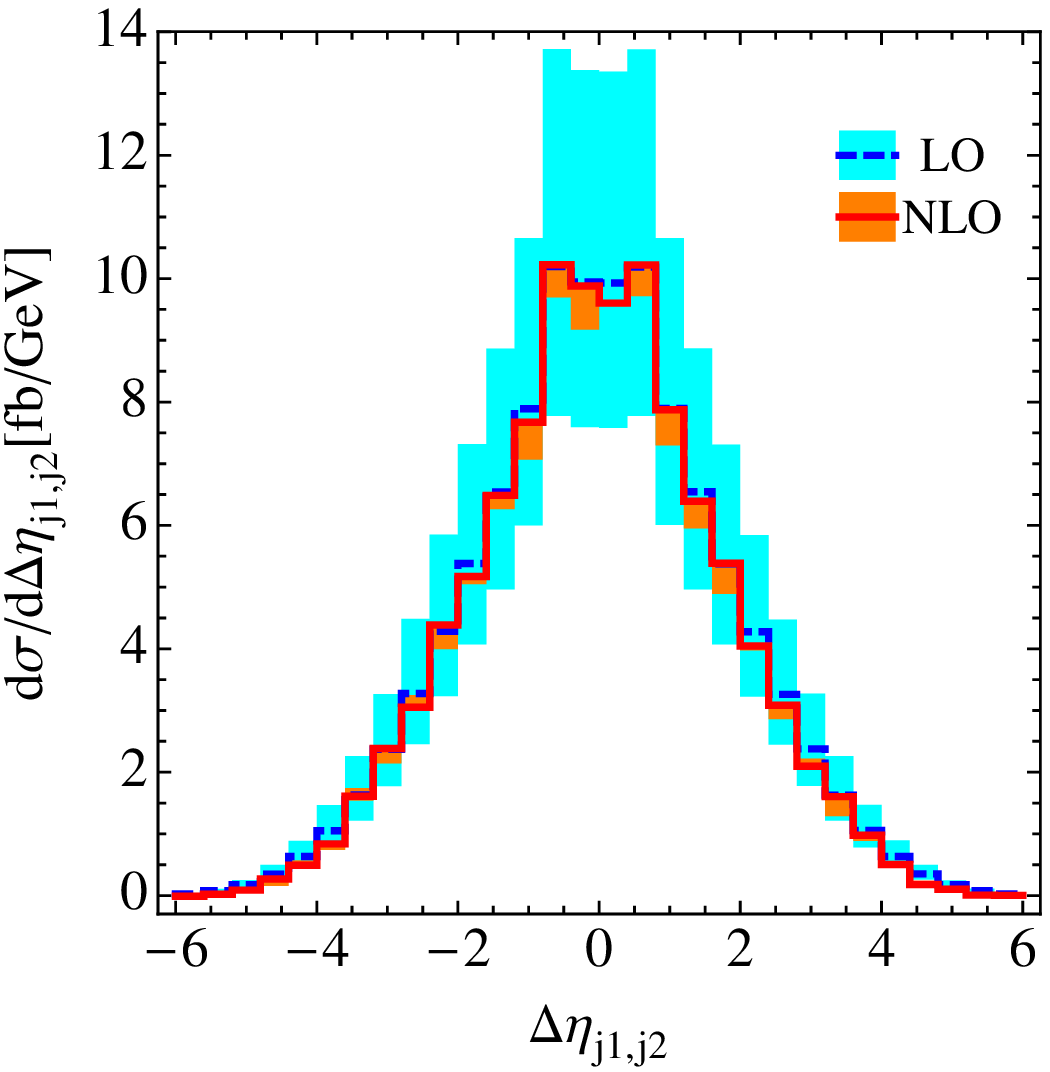}
\caption{Distributions of lepton opening angle and jet pseudorapidity difference for the process $\nobreak{pp\to(W^+\to\nu_ee^+)(W^-\to \mu^-\bar{\nu}_\mu)jj}$ at the $7~$TeV LHC. LO results
are shown in blue, NLO results in red. The uncertainty bands are for scale $m_W<\mu<4m_W$ and the solid lines show the results at $\mu=2m_W$. }
\label{kinhiggs}
\end{center}
\end{figure}

Finally I present two kinematic distributions for this process which are relevant for a Higgs boson search at the LHC. The left pane of figure~\ref{kinhiggs} plots the 
relative azimuthal angle between the leptons which peaks at $\phi_{e^-\mu^+}=\pi$. This is in contrast to leptons produced via the mechanism
$H\to WW\to e^+ \mu^- \nu\nu$ where this angle tends to be small. The pseudorapidity difference between the two leading jets, $\Delta\eta_{j1j2}=\eta_{j1}-\eta_{j2}$, is plotted in the right pane of figure~\ref{kinhiggs}. This is a useful distribution for studying Higgs boson production via WBF - this mechanism leads to jets
which tend to have a large $|\Delta\eta_{j1j2}|$. For a Higgs produced via gluon fusion and, as we see here for $\pptopm$, this distribution is peaked 
around $|\Delta\eta_{j1j2}|=0$. The significant reduction in 
theoretical scale uncertainties can also be seen in these distributions, and there is no observed shape change in going from LO to NLO. These observations were
typical of all kinematic distributions considered in \cite{Melia:2011dw}.

\section{Conclusion}
 In this talk I have presented the NLO QCD corrections for the process $\pptopp$ and the process $\pptopm$ which were computed using the method of 
$D$-dimensional generalised unitarity. A significant reduction in the theoretical uncertainties of an LHC prediction is observed for both processes.
The process $\pptopp$ has been implemented in the {\tt POWHEG BOX} which matches the NLO result with a parton shower. I look forward to measurements
of pairs of weak bosons and jets at the LHC.
\newline

\section*{Acknowledgements}
\noindent I wish to thank the organisers of RADCOR2011 for a really fantastic conference and for providing financial support. This talk is based on
work done in collaboration with Kirill Melnikov, Paolo Nason, Raoul R\"ontsch, and Giulia Zanderighi and draws on the papers 
\cite{Melia:2010bm,Melia:2011gk,Melia:2011dw}. This research is supported by the British Science and Technology Facilities Council.

\end{document}